\documentclass[aps,amsfonts,amsmath,prd,preprint,nofootinbib,tightenlines]{revtex4}
\usepackage{epsf}

\newcommand{\beq}{\begin{equation}}
\newcommand{\eeq}{\end{equation}}

\def\be{\begin{equation}}
\def\ee{\end{equation}}
\def\beqa{\begin{eqnarray}}
\def\eeqa{\end{eqnarray}}

\begin{document}

\title{Probabilities in the Arkani-Hamed-Dimopolous-Kachru landscape}
\author{Delia Schwartz-Perlov}
\affiliation{Institute of Cosmology, Department of Physics and
Astronomy\\
Tufts University, Medford, MA 02155, USA }

\begin{abstract}
In a previous paper we found that in the context of the string
theory ``discretuum'' proposed by Bousso and Polchinski, the
cosmological constant probability distribution varies wildly.
However, the successful anthropic predictions of the cosmological
constant depend crucially on the assumption of a flat prior
distribution. We conjectured that the staggered character of our
Bousso-Polchinski distribution will arise in any landscape model
which generates a dense spectrum of low-energy constants from a wide
distribution of states in the parameter space of the fundamental
theory.  Here we calculate the volume distribution for $\Lambda$ in
the simpler Arkani-Hamed-Dimopolous-Kachru landscape model, and
indeed this conjecture is borne out.

\end{abstract}
\maketitle

\section{Introduction}
While inflationary cosmologists \cite{AV83,Linde86,Starobinsky} have
long since realized that inflation generically gives rise to a
multiverse, much more recently string theorists have arrived at a
complementary world view \cite{BP,Susskind,duff}. Despite a quest to
uncover a single unique solution to the Laws of Nature, it seems as
though string theory admits a vast array of possible solutions. Each
solution, or vacuum state, represents a possible type of bubble
universe, governed by its own low-energy laws of physics.

\begin{figure}
\begin{center}
\leavevmode\epsfxsize=5in\epsfbox{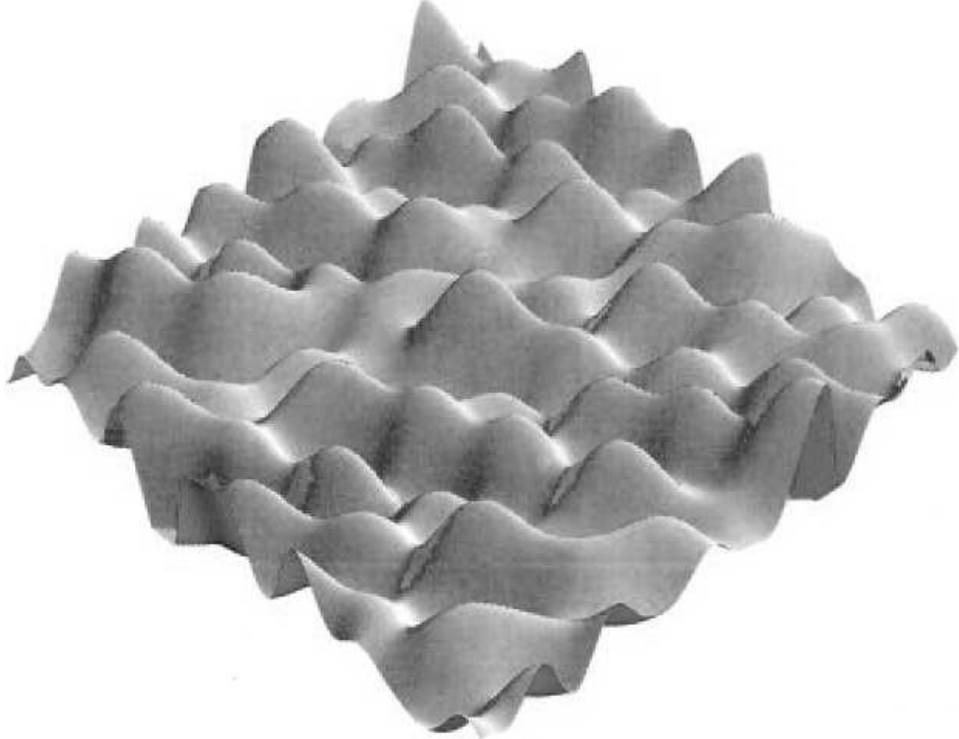}
\end{center}
\caption{ The string theory landscape. The vertical axis represents
the value of potential energy density (or equivalently, the
cosmological constant),
 and the two horizontal directions represent two out of hundreds of directions in parameter space,
 which characterize each vacuum solution.  Each valley represents a metastable vacuum solution which may have
 negative, zero or positive cosmological constant.} \label{landscape}
\end{figure}

One can depict each string theory vacuum solution as a local minimum
in a multidimensional potential energy diagram known as the string
theory landscape as illustrated in Fig.\ \ref{landscape}.  This
landscape of possibilities is expected to have many high-energy
metastable false vacua which can decay through bubble
nucleation\cite{CdL,Parke,BT}. Bubbles of lower-energy vacuum can
nucleate and expand in the high-energy vacuum background. If the
``daughter'' vacuum has a positive energy density, then inverse
transitions are also possible, allowing bubbles of high-energy
vacuum to nucleate within low-energy vacua
\cite{EWeinberg,recycling}. But if the ``daughter'' vacuum has
negative or zero-energy, recycling cannot take place.  We will call
vacua from which new bubbles can nucleate non-terminal, or
recyclable vacua, while those which do not recycle will be called
terminal vacua. This recycling process will populate the multiverse
with bubbles nested within bubbles of each and every possible type.

Most of these bubbles will never be hospitable to life. For example,
bubbles with large positive cosmological constant do not allow for
structures such as galaxies or atoms to form
\cite{Weinberg87,Linde87}. And bubbles with large negative
cosmological constant collapse long before life has a chance to
evolve.  However, because the landscape of possibilities is so
large, there will also be many bubbles which do provide a suitable
environment for life to flourish.  Obviously we live in one of these
``friendly'' bubbles.

In the framework of the multiverse, some physical parameters that
were once thought of as fundamental universal parameters get demoted
to local environmental parameters.  We no longer expect to calculate
the ``constants'' from first principles. Instead we are compelled to
calculate how the ``constants'' are distributed throughout the
multiverse.  If we assume we are a typical civilization, we should
expect to observe values near the peak of the distribution
\cite{AV95}.

The multiverse paradigm has led to the successful so-called
anthropic prediction of the cosmological constant
\cite{Weinberg87,Linde87,AV95,Efstathiou,MSW,GLV,Bludman,AV05}.
Theoretically we expect the magnitude of the cosmological
constant\footnote{Throughout this paper we use reduced Planck units,
$M_p^2/8\pi =1$, where $M_p$ is the Planck mass.} $\Lambda \sim 1$,
but the observed value is $\Lambda_0 \sim 10^{-120}$.  This has been
one of the biggest problems in theoretical
physics\footnote{Furthermore, $\Lambda_0\sim\rho_{m0}$ where
$\rho_{m0}$ is the present matter density.  The smallness of
$\Lambda$, and the fact that it happens to coincide with the present
matter density are collectively known as the cosmological constant
problems.}.

The probability for a randomly picked observer to measure a given
value of $\Lambda$ can be expressed as \cite{AV95} \be
P_{\text{obs}}(\Lambda)\propto P(\Lambda)n_{\text{obs}}(\Lambda),
\label{Pobs} \ee where $P(\Lambda)$ is the volume fraction of
regions with a given value of $\Lambda$ and
$n_{\text{obs}}(\Lambda)$ is the number of observers per unit
thermalized volume.

The factor $n_{\text{obs}}(\Lambda)$ takes into account selection
effects and is sometimes called the anthropic factor. It is in
general very difficult to calculate. However, it has been shown
\cite{Weinberg87} that the function $n_{\text{obs}}(\Lambda)$ is
only substantially different from zero in a tiny window of width
\beq \Delta\Lambda_A\sim 100\Lambda_0 \sim 10^{-118}
\label{DeltaLambdaA} \eeq around $\Lambda =0$. $\Delta\Lambda_A$ is
sometimes called the Weinberg window or the anthropic range.

The volume factor $P(\Lambda)$ depends on the dynamics of eternal
inflation and on the underlying fundamental theory. However, it has
been argued \cite{AV96,Weinberg96} that it should be accurately
approximated by a flat distribution, \be P(\Lambda)\approx {\rm
const} \label{flat} \ee because the anthropic range
(\ref{DeltaLambdaA}) is vastly less than the expected Planck scale
range of variation of $\Lambda$. A smooth function varying on this
large characteristic scale will be nearly constant within the minute
anthropic interval.

A tiny non-zero value for $\Lambda$ was predicted
\cite{Weinberg87,Linde87,AV95,BP} when the theoretical vogue was to
believe that a deep symmetry forced the cosmological constant to be
zero. One should keep in mind, however, that the successful
anthropic prediction for $\Lambda$ depends critically on the
assumption of a flat volume distribution (\ref{flat}).

In \cite{SPV} we used the new prescription introduced in
\cite{GSPVW} to calculate bubble abundances in an eternally
inflating spacetime to actually calculate the volume distribution
for the cosmological constant $\Lambda$ in the context of the
Bousso-Polchinski (hereafter BP) landscape model. We found that the
resulting distribution has a staggered appearance, in conflict with
the heuristically expected flat distribution.

One might think that the staggered distribution is a feature of the
BP model.  However, in this paper we calculate the volume
distribution for $\Lambda$ in a simpler landscape model proposed by
Arkani-Hamed, Dimopolous and Kachru \cite{AHDK} (hereafter called
the ADK model) and once again find a wildly varying distribution for
$\Lambda$.

%In the ADK landscape $J$ independent scalar fields $\phi_a$,
%$a=1,...,J$, contribute to the cosmological constant.  Each field
%$\phi_a$ can be in one of two minima $V_{a+}$ and $V_{a-}$, see
%Fig.\ \ref{potential}. This theory represents a landscape of $2^J$
%vacua labeled by $\{\eta\}=\{\eta_1,...,\eta_J\}$ with $\eta_a= \pm
%1$ indicating whether the $a'th$ field is in minima $V_{a+}$ or
%$V_{a-}$.

%The cosmological constant is given by \be \Lambda_{\{\eta\}} =
%\bar{\Lambda}+\sum_{a=1}^{J} \eta_a V_{{diff}~a}\ee where $\bar
%{\Lambda}$ is the average value of $\Lambda$ for the spectrum and
%$V_{{diff}~a}$ is related to the change in $\Lambda$ when the field
%configuration changes from one minimum to another in the $a'th$
%direction.

%The field configuration in an inflating region can change when one
%of the fields $\phi_a$ tunnels to it's other minimum through bubble
%nucleation. As in the BP model, bubbles nucleate within bubbles,
%allowing the universe to start off with an arbitrary large
%cosmological constant, and then to diffuse through the ADK landscape
%of possible vacua.

The outline of this paper is as follows: In section \ref{adkmodel}
we will describe the ADK landscape model. In section
\ref{probabilities} we use the new prescription \cite{GSPVW} for
calculating probabilities to calculate the volume distribution for
$\Lambda$ in the ADK model. We will conclude with a discussion in
section \ref{conclusions}.

\section{The ADK landscape}\label{adkmodel}

\begin{figure}
\begin{center}
\leavevmode\epsfxsize=5in\epsfbox{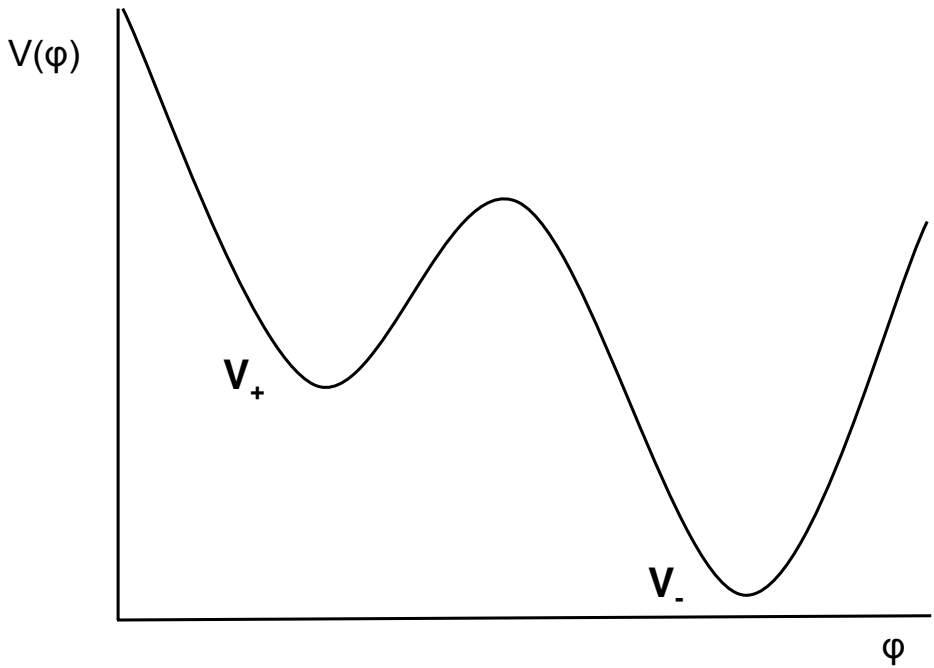}
\end{center}
\caption{ Potential with two metastable minima.} \label{potential}
\end{figure}

Following Ref.~\cite{AHDK} we consider a single scalar field $\phi$
with a general quartic potential.  Also we assume that the theory
has two minima at $\phi_{\pm}$ with vacuum energies $V_{\pm}$, and
we take $V_+ \geq V_-$.  Thus $V_+$ ($V_-$) represents the energy of
the false vacuum (true vacuum) at $\phi_+$ ($\phi_-$)(see Fig.\
\ref{potential}).  We can label the vacua by $\eta=\pm1$, and define
\be V_{\eta}=V_{av}+\eta V_{\text{dif}}\ee with \be
V_{av}={\frac{1}{2}}(V_++V_-)\ee and \be
V_{\text{dif}}={\frac{1}{2}}(V_+-V_-)\ee

Now consider a theory with J scalar fields $\phi_a$, $a=1,...,J$,
each having independent quartic potentials $V_a(\phi_a)$ such that
the potential is the sum of $J$ independent potentials \be V=
\sum_{a=1}^{J} V_a(\phi_a)\ee This theory represents a landscape of
$2^J$ vacua labeled by $\{\eta\}=\{\eta_1,...,\eta_J\}$ with
$\eta_a= \pm 1$.

The cosmological constant is given by \be\Lambda_{\{\eta\}} =
\bar{\Lambda}+\sum_{a=1}^{J} \eta_a V_{{\text{dif}}~a}\ee where
\be\bar {\Lambda}=\sum_{a=1}^{J} V_{{av~}a}=J \bar {V}_{av~}\ee

If we consider a specific bubble which is completely specified by
$\{\eta_1,...,\eta_a,...\eta_J\}$, the field configuration in an
inflating region can change to $\{\eta_1,...,\eta_a \pm
2,...\eta_J\}$ when one of the fields $\phi_a$ tunnels to its other
minimum through bubble nucleation. Thus, as in the BP model, the
universe can start off with an arbitrary large cosmological
constant, and then diffuse through the ADK landscape of possible
vacua as bubbles nucleate one within the other.

ADK asked the question, does a landscape with $2^J$ vacua guarantee
that we can solve the cosmological constant problem? In the ADK
model the histogram of the number of vacua per bin of $\Lambda$ is
well approximated by a Gaussian distribution because $\Lambda$ is
the sum of many independent components. If $\Lambda =0$ is on the
tail of the Gaussian the cosmological constant will not scan around
$0$, but if it is near the peak we can expect to find a dense enough
spectrum of vacua to account for $\Lambda_0$. Thus ADK concluded
that for the landscape to solve the cosmological constant problem,
either a tiny cosmological constant can accidentally arise on the
tail of the Gaussian\footnote{Of course if this were the case then
the anthropic explanation for $\Lambda_0$ would not be applicable -
we simply land up at $\Lambda_0$ accidentally.} or the Gaussian must
be peaked close enough to $\Lambda=0$.

We wish to point out that even if the Gaussian is peaked around
$\Lambda=0$, yielding a sufficiently dense spectrum of $\Lambda$ in
the anthropic range $\Delta\Lambda_A$, this is still not good enough
to validate the anthropic resolution of the cosmological constant
problems.  It could be that the probabilities of these anthropic
vacua differ dramatically (we have learned from our calculation in
the BP model \cite{SPV} that the probabilities tend to span many
orders of magnitude) with one or two dominating the distribution.

%and we may need a huge number of vacua in $\Delta\Lambda_A$ to
%justify the anthropic explanation for $\Lambda_0$.

We will now calculate the volume distribution $P(\Lambda)$ for the
ADK model.

\section{Probabilities in the ADK landscape}\label{probabilities}
In this section we study the volume distribution for $\Lambda$ in
the ADK landscape using the prescription of \cite{GSPVW} which we
outline below. We expect to find a wildly varying distribution like
the one found for the BP model studied in \cite{SPV}.

\subsection{Summary of probability prescription}
Suppose we have a theory with a discrete set of vacua, labeled by
index $j$, and having cosmological constants $\Lambda_j$. The volume
distribution is given by \cite{GSPVW} \be P_j \propto p_j Z_j^3,
\label{PpZ} \ee where $p_j$ is the relative abundance of bubbles of
type $j$ and $Z_j$ is (roughly) the amount of slow-roll inflationary
expansion inside the bubble after nucleation (so that $Z_j^3$ is the
volume slow-roll expansion factor).

The bubble abundances $p_j$ can be related to the comoving volume
fractions $f_j(t)$ which obey the evolution equation
\cite{recycling} \beq {df_j\over{dt}}=\sum_i (-\kappa_{ij}f_j +
\kappa_{ji}f_i), \label{dfdt} \eeq where the first term on the
right-hand side accounts for loss of comoving volume due to bubbles
of type $i$ nucleating within those of type $j$, and the second term
reflects the increase of comoving volume due to nucleation of
type-$j$ bubbles within type-$i$ bubbles.

The transition rate $\kappa_{ij}$ is defined as the probability per
unit time for an observer who is currently in vacuum $j$ to find
herself in vacuum $i$ and is given by \beq
\kappa_{ij}=\Gamma_{ij}{4\pi\over{3}}H_j^{-4}, \label{kappa} \eeq
where $\Gamma_{ij}$ is the bubble nucleation rate per unit physical
spacetime volume (same as $\lambda_{ij}$ in \cite{GSPVW}) and \be
H_j = (\Lambda_j/3)^{1/2} \label{Hj} \ee is the expansion rate in
vacuum $j$.

Eq.~(\ref{dfdt}) can be written in a vector form, \beq {d{\bf
f}\over{dt}}={\mathbf M}{\bf f}, \label{matrixf} \eeq where ${\bf
f(t)}\equiv \{ f_j(t)\}$ and \beq
M_{ij}=\kappa_{ij}-\delta_{ij}\sum_r \kappa_{ri}. \label{Mij} \eeq

Let's label the eigenvalue (of the transition matrix ${\mathbf M}$)
which has the smallest negative real part (in magnitude), $-q$, and
the corresponding eigenvector, $\mathbf{s}$.  Then it can be shown
that the bubble abundances $p_j$ are given by \beq p_j\propto
\sum_\alpha H_\alpha^q \kappa_{j\alpha} s_\alpha. \label{pJaume}
\eeq where the summation is over all recyclable vacua which can
directly tunnel to $j$.

The problem of calculating $p_j$ has thus been reduced to finding
the dominant eigenvalue $q$ and the corresponding eigenvector ${\bf
s}$ of the transition matrix $\mathbf{M}$.  To calculate the
elements of $\mathbf{M}$ we need to calculate the bubble nucleation
rates specific to the landscape model we are studying.

\subsection{Nucleation rates in the ADK landscape} Transitions
between neighboring vacua, which change one of the integers $\eta_a$
by $\pm 2$ can occur through bubble nucleation. The bubbles are
bounded by thin branes, with tension $\tau_a$.  By analogy with the
BP model\footnote{More specifically for the case of $n_a=1$ where
$n_a$ is a flux quantum in the BP model.} we will take the brane
tension in the ADK model to be \be\tau_a \equiv
\sqrt{\Delta\Lambda_a}= \sqrt{2V_{\text{dif},a}}\label{ADKtauj}\ee

Transitions with multiple brane nucleation, in which several
$\eta_a$ are changed at once, are likely to be strongly suppressed
\cite{Megevand}, and we shall disregard them here.

%As in the BP model, we distinguish between the recyclable,
%non-terminal vacua, with $\Lambda_j>0$, and the non-recyclable,
%``terminal vacua", for which $\Lambda_j\leq 0$.

The bubble nucleation rate $\Gamma_{ij}$ per unit spacetime volume
can be expressed as \cite{CdL} \be \Gamma_{ij}=A_{ij} \exp^{-B_{ij}}
\label{Gamma} \ee with \beq B_{ij}=I_{ij}-S_j \label{Bij} \eeq Here,
$I_{ij}$ is the Coleman-DeLuccia instanton action and \beq
S_j=-{8\pi^2\over{H_j^2}} \label{Sj} \eeq is the background
Euclidean action of de Sitter space with expansion rate \be
H_j=\sqrt{\Lambda_j/3}.\label{hubble}\ee

In the relevant case of a thin-wall bubble, the instanton action
$I_{ij}$ has been calculated in Refs.~\cite{CdL,BT}. It depends on
the values of $\Lambda$ inside and outside the bubble and on the
brane tension $\tau$.

Let us first consider a bubble which changes one $\eta_a$ from
$\eta_a=+1$ to $\eta_a=-1$.  The resulting change in the
cosmological constant is given by \be
|\Delta\Lambda_a|=2V_{\text{dif},a} \label{ADKDeltaLambda} \ee and
the exponent in the tunneling rate (\ref{Gamma}) can be expressed as
\be B_{a\downarrow} = B_{a\downarrow}^{flatspace} r(x,y).
\label{ADKBdown} \ee $B_{a\downarrow}^{flatspace}$ is the flat space
bounce action, \be B_{a\downarrow}^{flatspace}= \frac{27
\pi^2}{2}\frac{\tau_a^4}{|\Delta \Lambda_a|^3}. \ee With the aid of
Eqs.~(\ref{ADKtauj}),(\ref{ADKDeltaLambda}) it can be expressed as
\be B_{a\downarrow}^{flatspace}= \frac{27 \pi^2}{4V_{\text{dif},a}}
\label{ADKBflat} \ee

The gravitational correction factor $r(x,y)$ is given by
\cite{Parke} \be r(x,y) = \frac{2[(1+x
y)-(1+2xy+x^2)^{\frac{1}{2}}]}{x^2(y^2-1)(1+2xy+x^2)^{\frac{1}{2}}}
\label{gravfactor} \ee  with the dimensionless parameters \be
x\equiv \frac{3\tau_a^2}{4|\Delta\Lambda_a|}=\frac{3}{4}
\label{ADKx}\ee and \be y\equiv
\frac{2\Lambda}{|\Delta\Lambda_a|}-1, \label{ADKy} \ee where
$\Lambda$ is the background value prior to nucleation.
%\frac{\Lambda_{n_j}+\Lambda_{n_j-1}}{\Lambda_{n_j}-\Lambda_{n_j-1}}=
%\frac{\Lambda_{n_j}+\Lambda_{n_j-1}}{\Delta \Lambda}.
%\ee

The prefactors $A_{ij}$ in (\ref{Gamma}) can be estimated as
\cite{Jaume} \beq A_{ij} \sim 1 \eeq

If the vacuum $\{\eta_1...\eta_{a-1},\eta_a-2,\eta_{a+1}....\}$
still has a positive energy density, then an upward transition from
$\{\eta_1...\eta_{a-1},\eta_a-2,\eta_{a+1}....\}$ to
$\{\eta_1...\eta_{a-1},\eta_a,\eta_{a+1}....\}$ is also possible.
The corresponding transition rate is characterized by the same
instanton action and the same prefactor \cite{EWeinberg}, and it
follows from Eqs. (\ref{Gamma}), (\ref{Bij}) and (\ref{hubble}) that
the upward and downward nucleation rates are related by \be
\Gamma_{ji} = \Gamma_{ij} \exp\left[-24 \pi^2
\left(\frac{1}{\Lambda_{i}}-\frac{1}{\Lambda_{j}}\right)\right]
\label{updown} \ee where $\Lambda_j>\Lambda_i$.  As expected the
transition rate from $\eta_a=-1$ up to $\eta_{a}=+1$ is suppressed
relative to that from $\eta_{a}=+1$ down to $\eta_{a}=-1$. The
closer we are to $\Lambda_i=0$, the more suppressed are the upward
transitions $i\to j$ relative to the downward ones.

We will now investigate the dependence of the tunneling exponent
$B_{a\downarrow}$ on the parameters of the model in the limits of
small and large $\Lambda$. For $\Lambda\ll |\Delta\Lambda_a|$, we
have $y\approx -1$, and Eq.\ (\ref{gravfactor}) gives \beq r(y\to
-1)=16. \eeq The inclusion of gravity increases the tunneling
exponent causing a suppression of the nucleation rate.

In the opposite limit, $\Lambda\gg|\Delta\Lambda_a|$, $y\gg 1$, \beq
r(y\gg 1)\approx \sqrt{2}(xy)^{-3/2}, \eeq and \beq
B_{a\downarrow}\approx 27\pi^2 \sqrt{V_{\text{dif},a}}
\left(\frac{2}{3\Lambda}\right)^{3/2}. \label{ADKLargeLambda} \eeq

%The gravitational factor $r$ for the ADK model  as a function of
%$\Lambda/|\Delta\Lambda_a|$ (assuming $x=3/4$) is shown in Fig.\
%\ref{adkrvsz}.
For large values of $\Lambda$, $r\ll 1$, so the nucleation rate is
enhanced. The tunneling action must always be large enough to
justify the use of the semi-classical approximation:
$B_{a\downarrow}\gg 1$, or \be \Lambda \ll 30
V_{\text{dif},a}^{1/3}\label{ADKsemiclasslimit}.\eeq

If $V_{\text{dif},a}$ and $\Lambda$ are changed simultaneously,
keeping the ratios $\Lambda/V_{\text{dif},a}$ fixed, then $x$ and
$y$ do not change, and it is clear from
Eqs.~(\ref{ADKBdown}),(\ref{ADKBflat}),(\ref{gravfactor}) that the
nucleation exponents scale as $B_{ij}\propto \Lambda^{-1}$.  This
shows that bubble nucleation rates are strongly suppressed when the
energy scales of $V_{\text{dif},a}$ and $\Lambda$ are well below the
Planck scale.

\subsection{Bubble abundances in the ADK model} 
\begin{figure}
\begin{center}
\leavevmode\epsfxsize=5in\epsfbox{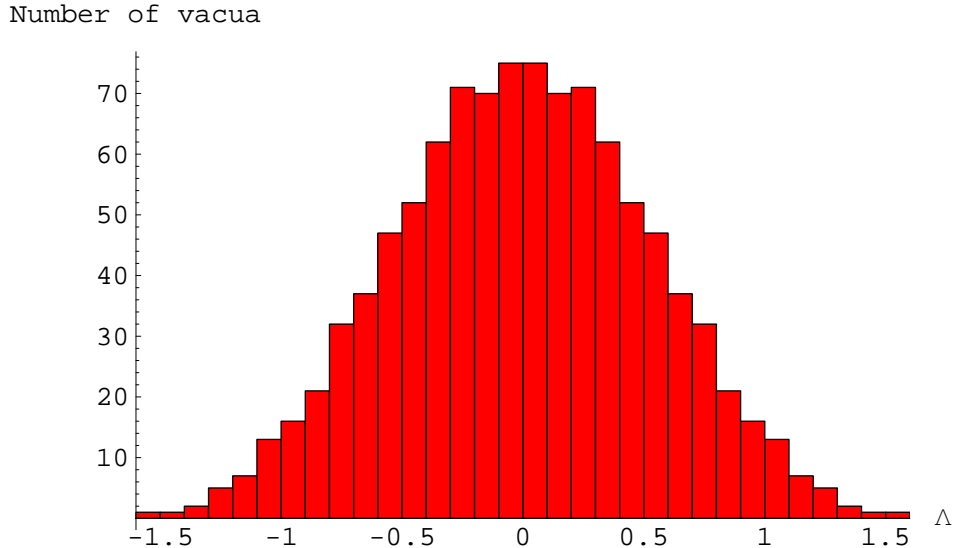}
\end{center}
\caption{The spectrum of vacua for a $J=10$ ADK landscape with
parameters given in Eq.\ (\ref{ADKparam})} \label{ADKhistogram}
\end{figure}

We will calculate the
bubble abundances for a $J=10$ ADK model, containing $2^{10}$ vacua,
with parameter values: \be V_{\text{dif},a}=\{ 0.0514, 0.0814,
0.0885, 0.1081, 0.1378, 0.1475,
    0.1790, 0.2226, 0.2467, 0.2523\}\label{ADKparam}\ee
and $\bar{\Lambda}=0$. A histogram of the number of vacua vs.
$\Lambda$ for this model is given in Fig.\ \ref{ADKhistogram}.

In direct analogy with our calculation of probabilities for the BP
landscape \cite{SPV}, we resort to perturbative techniques, where we
use the smallness of upward transitions as a small parameter. The
slowest vacuum to decay was singled out as the dominant vacuum
$\alpha_*$. To zero'th order in perturbation theory (hereafter PT)
the only vacua which aquire non-zero probabilities are those that
are direct offspring from the dominant vacuum $\alpha_*$.  These
vacua will have large negative cosmological constants.

\begin{figure}
\begin{center}
\leavevmode\epsfxsize=5in\epsfbox{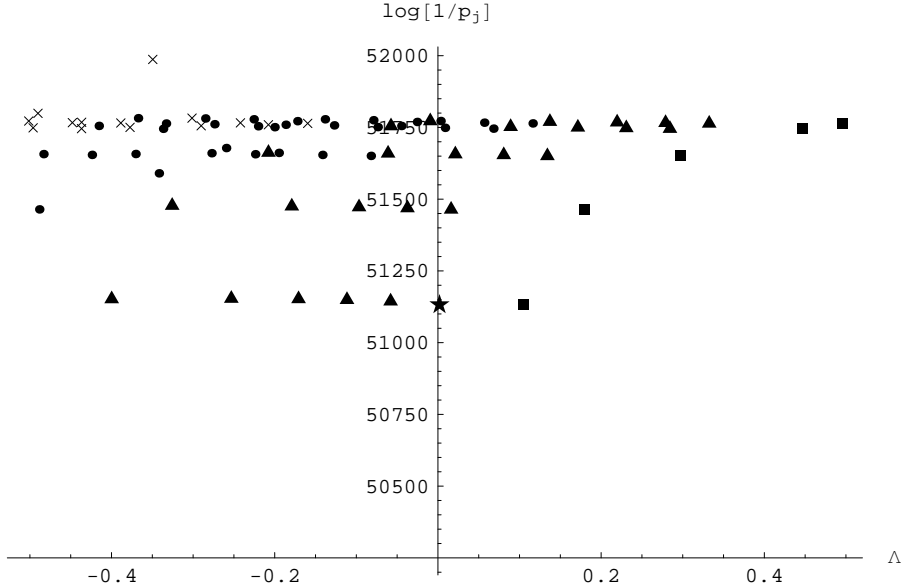}
\end{center}
\caption{Plot of $\log_{10}(1/p_j)$ vs. $\Lambda_j$ for the ADK
model with parameters given in (\ref{ADKparam}) and $x=3/4$ (see
(\ref{ADKx})). The star marks the dominant vacuum $\alpha_*$.}
\label{ADKprobabilities1}
\end{figure}

The results for the first order bubble abundance factors $p_j$ are
shown in Fig.\ref{ADKprobabilities1}.

%\begin{figure}
%\begin{center}
%\leavevmode\epsfxsize=5in\epsfbox{ADKprobabilities2.eps}
%\end{center}
%\caption{Plot of $\log_{10}(1/p_j)$ vs. $\Lambda_j$ for the ADK
%model with parameters given in (\ref{ADKparam}) and $x=3/16$. The
%star marks the dominant vacuum $\alpha_*$.}
%\label{ADKprobabilities2}
%\end{figure}

The dominant site in Fig.~\ref{ADKprobabilities1}  has
``coordinates" (-1, 1, -1, 1, 1, -1, 1, -1, -1, 1) and has a very
small\footnote{Small relative to the values in the spectrum of our
toy model.} cosmological constant $\Lambda_*=0.0019$. The five
squares represent the vacua which can be reached from $\alpha_*$ via
one upward jump. Their bubble abundances are so low because
$\Lambda_*$ is so small resulting in very suppressed upward jumps
(see Eq. (\ref{updown})). Each site represented by a square can then
jump down in 5 ways (excluding jumps back to the dominant site
itself) to the sites depicted as triangles which can in turn jump
down to the circles followed by crosses.

The probabilities $p_j$ shown here are more suppressed than the BP
results in \cite{SPV} because $\Lambda_*$ happens to be much
smaller. This is simply a consequence of the different parameters.
Also, unlike the BP results, it appears as though vacua which result
from downward jumps from a given up jump, have almost the same
probability. This ``flatness'' is fictitious - there are actually a
few orders of magnitude difference amongst these vacua which is hard
to see graphically because of the scale. However, overall the
``staggered'' nature of the distribution is evident and the
similarity to the BP model is clear.

In addition to the bubble abundance factor $p_j$, the volume
distribution (\ref{PpZ}) includes the slow-roll expansion factor
$Z_j$.  There is no reason to expect the expansion factor to tame
the wildly varying bubble abundance distribution, and thus we
conclude that the volume distribution for $\Lambda$ will have the
same form as that calculated for the bubble abundances.

\section{Conclusions and Discussion}\label{conclusions}

In \cite{SPV} we found that for the Bousso-Polchinski string theory
landscape, the cosmological constant probability distribution varies
dramatically for vacua which have close values of $\Lambda$.  We
have shown that this behavior persists in the case of the
Arkani-Hamed-Dimopolous-Kachru landscape model.

This result was expected.  Our probability prescription picks a
dominant vacuum and all other vacua are reached from it via a
sequence of upward and downward jumps. To zero'th order in PT only
the progeny of the dominant vacuum have non-zero probabilities.
These vacua were shown to have large\footnote{Large compared to the
size of $\Lambda_0$.} negative $\Lambda$. To the first order in PT
only a small subset of vacua related to the dominant site via one
upward jump and any number of downward jumps gain non-zero
probabilities. The probabilities of these vacua are proportional to
the tunneling transition rates of the jumps. The tunneling
transition rates have an exponential dependence on the parameters of
the theory and consequently the probabilities span many orders of
magnitude, in both landscape models considered.

Furthermore, it is exceedingly unlikely that one of these first
order vacua should be in the anthropic range - we just don't have
enough of them. So what happens if we go to second order in PT?

Calculations indicate that vacua which can be reached via two upward
jumps and subsequent downward jumps would gain some tiny
probabilities.  There would be many more vacua which can be reached
via paths including two upward jumps instead of only one, but we
would still need to consider higher and higher orders of PT before a
sufficient fraction of the theory's vacua can be infused with
probability. Going to higher orders is technically prohibitive.

Although the distributions we have calculated do not give a flat
distribution to first order, we cannot conclude that the anthropic
prediction of the cosmological constant was a fluke.  The
distributions we are able to calculate are simply the tip of the
iceberg.  It is still entirely possible that vacua in the anthropic
range are smoothly distributed.

So how do we proceed from here?  How do we look beyond the first
order results we have found for a tiny subset of vacua in our
landscape? Currently we do not have a definitive answer to this
question.  But work is underway to try to elucidate what essential
features of a given landscape model will guarantee that (once we
have many vacua in the anthropic range) a sufficient number of the
most probable anthropic vacua will have close enough probabilities
to ensure that the distribution can be considered to be smooth.

\section{Acknowledgements}
I would like to thank Ken Olum and Alex Vilenkin for their support
and also for many useful suggestions for this manuscript.

%\textbf{Discuss: How our prescription picks one dom and then the
%probabilities get spread out - upward jumps are very sensitive to
%$\Lambda$. Multiple downward jumps give rise to many vacua -
%combinatorics comes in.  Overall factor of $E^{(-24
%\pi^2/\Lambda_*)}$ and then relative enhancement if you didn't jump
%too high.  What would be expected if we could jump up several times
%etc. Perhaps give recursive relation for $s_{\alpha}'s$ and corr.
%$p's$}

\end{document}